\documentclass[12pt]{article}
\usepackage{graphicx}
\def\beq{\begin{equation}}
\def\eeq{\end{equation}}
\def\bea{\begin{eqnarray}}
\def\eea{\end{eqnarray}}
\def\nn{\nonumber}

\def\sss{\scriptscriptstyle}

\def\roughly#1{\mathrel{\raise.3ex\hbox
{$#1$\kern-.75em\lower1ex\hbox{$\sim$}}}}
\def\lsim{\roughly<}

\def\sla#1{\raise.15ex\hbox{$/$}\kern-.57em #1}% Feynman slash
\def\bra#1{\left\langle #1\right|}
\def\ket#1{\left| #1\right\rangle}
\def\ks{K_{\sss S}}

\def\bd{B_d^0}

\def\bdbar{{\bar B}_d^0}
\def\btos{{\bar b} \to {\bar s}}
\def\ANPq{{\cal A}^q}

\def\ApNPqph{{\cal A}^{\prime,q} e^{i \Phi'_q}}
\def\ApNPCqph{{\cal A}^{\prime {\sss C}, q} e^{i \Phi_q^{\prime C}}}
\def\ApNPCuph{{\cal A}^{\prime {\sss C}, u} e^{i \Phi_u^{\prime C}}}
\def\ApNPCdph{{\cal A}^{\prime {\sss C}, d} e^{i \Phi_d^{\prime C}}}
\def\pewcp{P_{\sss EW}^{\prime\sss C}}
\def\pewp{P'_{\sss EW}}
\def\pewcpnp{P_{\sss EW, NP}^{\prime\sss C}}
\def\pewpnp{P'_{\sss EW, NP}}
\def\ApNPuph{{\cal A}^{\prime,u} e^{i \Phi'_u}}
\def\ApNPdph{{\cal A}^{\prime,d} e^{i \Phi'_d}}
\def\ApNPcomb{{\cal A}^{\prime, comb} e^{i \Phi'}}
\def\btopik{B \to \pi K}
\begin{document}

\begin{flushright}  
UdeM-GPP-TH-09-174 \\
\end{flushright}

\begin{center}
\bigskip
{\Large \bf \boldmath The $\btopik$ Puzzle: 2009 Update } \\
\bigskip
{\large Seungwon Baek $^{a,}$\footnote{sbaek1560@gmail.com},
Cheng-Wei Chiang $^{b,}$\footnote{chengwei@phy.ncu.edu.tw}
and David London $^{c,}$\footnote{london@lps.umontreal.ca}}
\\
\end{center}

\begin{flushleft}
~~~~~~$a$: {\it Institute of Basic Science and
Department of Physics, }\\
~~~~~~~~~~{\it Korea University, Seoul 136-701, Korea}\\
~~~~~~$b$: {\it Department of Physics and Center for
Mathematics and Theoretical Physics,}\\
~~~~~~~~~~{\it National Central University, Chungli,
Taiwan 320, Taiwan;}\\
~~~~~~~~~~{\it Institute of Physics, Academia Sinica,
Taipei, Taiwan 115, Taiwan}\\
~~~~~~$c$: {\it Physique des Particules, Universit\'e de
Montr\'eal,}\\
~~~~~~~~~~{\it C.P. 6128, succ. centre-ville,
Montr\'eal, QC, Canada H3C 3J7}
\end{flushleft}
\begin{center} 
%\bigskip (\today)
\vskip0.5cm
{\Large Abstract\\}
\vskip3truemm

\parbox[t]{\textwidth} {The measurements of $\btopik$ decays
have been in disagreement with the predictions of the
Standard Model (SM) for some time.  In this paper, we perform
an update of this puzzle using the latest (2008) data.  We
find that the situation has become far less clear.  A fit to
the $\btopik$ data alone suggests the presence of new physics
(NP).  Indeed, if one adds a constraint on the weak phase
$\gamma$ coming from independent measurements -- the SM fit
-- one finds that the fit is poor.  On the other hand, it is
not terrible.  If one is willing to accept some deficiencies
in the fit, it can be argued that the SM can explain the
$\btopik$ data. If one assumes NP, it is found to be present
only in the electroweak penguin amplitude, as before.
However, the fit is fair at best, and the improvement over
the SM is not particularly strong.  All and all, while the
$\btopik$ puzzle has not disappeared, it has become weaker.}
\end{center}

\thispagestyle{empty}
\newpage
\setcounter{page}{1}
% Decrease texheight (for preprint numbers) again
%\textheight 23.0 true cm
\baselineskip=14pt

The four $\btopik$ decays -- $B^+ \to \pi^+ K^0$ (designated
as $+0$ below), $B^+ \to \pi^0 K^+$ ($0+$), $\bd \to \pi^-
K^+$ ($-+$) and $\bd \to \pi^0 K^0$ ($00$) -- have evoked a
great deal of interest in recent years. There are nine
measurements of these processes that can be made: the four
branching ratios, the four direct CP asymmetries $A_{\sss
CP}$, and the mixing-induced CP asymmetry $S_{\sss CP}$ in
$\bd\to \pi^0K^0$.  There has been a continual disagreement
between this set of measurements and the predictions of the
standard model (SM); this discrepancy has been dubbed the
``$\btopik$ puzzle'' \cite{BFRS}.

The four $\btopik$ amplitudes, which obey a quadrilateral
isospin relation, can be written within the diagrammatic
approach \cite{GHLR}.  Here, the amplitudes are expressed in
terms of six diagrams\footnote{Note that we have neglected
the annihilation diagram $A'$, which is expected to be very
small in the SM.  In any case, its inclusion does not change
anything, since the diagrams can be redefined so that the
$\btopik$ amplitudes are still a function of six diagrams
\cite{Kim}.}: the color-favored and color-suppressed tree
amplitudes $T'$ and $C'$, the gluonic penguin amplitudes
$P'_{tc}$ and $P'_{uc}$, and the color-favored and
color-suppressed electroweak penguin amplitudes $\pewp$ and
$\pewcp$. (The primes on the amplitudes indicate $\btos$
transitions.) The amplitudes are given by
\bea
\label{fulldiagrams}
A^{+0} &\!\!=\!\!& -P'_{tc} + P'_{uc} e^{i\gamma} -\frac13
\pewcp ~, \nn\\
\sqrt{2} A^{0+} &\!\!=\!\!& -T' e^{i\gamma} -C' e^{i\gamma}
+P'_{tc} -~P'_{uc} e^{i\gamma} -~\pewp -\frac23 \pewcp ~,
\nn\\
A^{-+} &\!\!=\!\!& -T' e^{i\gamma} + P'_{tc} -P'_{uc}
e^{i\gamma} -\frac23 \pewcp ~, \nn\\
\sqrt{2} A^{00} &\!\!=\!\!& -C' e^{i\gamma} - P'_{tc} +P'_{uc}
e^{i\gamma} - \pewp -\frac13 \pewcp ~.
\eea
We have explicitly written the weak-phase dependence
(including the minus sign from $V_{tb}^* V_{ts}$ [in
$P'_{tc}$]), while the diagrams contain strong phases. The
amplitudes for the CP-conjugate processes can be obtained
from the above by changing the sign of the weak phase
($\gamma$).

Within the SM, to a good approximation, the diagrams $\pewp$
and $\pewcp$ can be related to $T'$ and $C'$ using flavor
SU(3) symmetry\footnote{Note that $\pewp$ and $\pewcp$ are
not written with a minus sign in Eq.~(\ref{fulldiagrams}),
despite containing the factor $V_{tb}^* V_{ts}$.  This sign
is included in the relations of these diagrams to $T'$ and
$C'$ [Eq.~(\ref{EWPrels})].} \cite{EWPs}:
\bea
\label{EWPrels}
\pewp & \!\!=\!\! & {3\over 4} {c_9 + c_{10} \over c_1 + c_2} R (T' +
C') \!+\!  {3\over 4} {c_9 - c_{10} \over c_1 - c_2} R (T' - C')
~, \nn\\
\pewcp & \!\!=\!\! & {3\over 4} {c_9 + c_{10} \over c_1 + c_2} R (T' +
C') \!-\!  {3\over 4} {c_9 - c_{10} \over c_1 - c_2} R (T' - C')
~.
\eea
Here, the $c_i$ are Wilson coefficients \cite{BuraseffH} and
$R \equiv \left\vert (V_{tb}^* V_{ts})/(V_{ub}^* V_{us})
\right\vert$.  In our fits we take $R=48.9 \pm 1.6$
\cite{CKMfitter}.

In Ref.~\cite{GHLR}, the relative sizes of the $\btopik$
diagrams were roughly estimated as
\beq
1 : |P'_{tc}| ~~,~~~~ {\cal O}({\bar\lambda}) : |T'|,~|\pewp|
~~,~~~~ {\cal O}({\bar\lambda}^2) : |C'|,~|P'_{uc}|,~|\pewcp|
~,
\label{ampsizes}
\eeq
where ${\bar\lambda} \sim 0.2$.  These estimates have since
been modified slightly.  First, in realistic models of QCD,
$|C'|$ is allowed to take somewhat larger values: $|C'/T'|
\lsim 0.6$ \cite{C'size}.  Second, it has been argued that
$|P'_{uc}|$ is actually a bit smaller, ${\cal
O}({\bar\lambda}^3)$ \cite{Kim}.

Before 2006, the $\btopik$ puzzle could be seen by comparing
two quantities which are expected to be approximately equal
in the SM.  However, with the data in 2006, it was no longer
possible to see such an effect.  It was therefore necessary
to perform a full fit to the data, and this was done in
Ref.~\cite{BL} (and before that, in Ref.~\cite{Baeketal}).
There it was found that a good fit could be obtained, but at
a serious cost.  If $P'_{uc}$ is excluded from the fit, then
$|C'/T'| = 1.6 \pm 0.3$ was required.  This is much larger
than the allowed value given above.  If $P'_{uc}$ is included
in the fit, a smaller value of $|C'/T'|=0.8 \pm 0.1$ was
obtained.  However, $|P'_{uc}/T'|= 1.7 \pm 0.6$ was found,
which is much larger than the expected value above.  In
addition, one found $\gamma = (30 \pm 7)^\circ$, which is
inconsistent with independent measurements.  In either case,
there was a clear indication that a 3-4$\sigma$ discrepancy
between the $\btopik$ data and the SM was present in 2006.

The 2008 data are shown in Table~\ref{tab:data}.  Compared to
2006, the only measurements which have changed significantly
are $A_{\sss CP}$ and $S_{\sss CP}$ in $\bd\to \pi^0K^0$.
However, as we will see, these changes have important
consequences.

\begin{table}[tbh]
\center
\begin{tabular}{cccc}
\hline
\hline
Mode & $BR[10^{-6}]$ & $A_{\sss CP}$ & $S_{\sss CP}$ \\ \hline
$B^+ \to \pi^+ K^0$ & $23.1 \pm 1.0$ & $0.009 \pm 0.025$ & \\
$B^+ \to \pi^0 K^+$ & $12.9 \pm 0.6$ & $0.050 \pm 0.025$ & \\
$\bd \to \pi^- K^+$ & $19.4 \pm 0.6$ & $-0.098^{+0.012}_{-0.011}$ & \\
$\bd \to \pi^0 K^0$ & $9.8 \pm 0.6$ & $-0.01 \pm 0.10$ &
$0.57 \pm 0.17$ \\
\hline
\hline
\end{tabular}
\caption{Branching ratios, direct CP asymmetries $A_{\sss
CP}$, and mixing-induced CP asymmetry $S_{\sss CP}$ (if
applicable) for the four $\btopik$ decay modes. The data are
taken from Refs.~\cite{HFAG} and \cite{piKrefs}.}
\label{tab:data}
\end{table}

In this paper, we explore the current status of the $\btopik$
puzzle by performing fits to the 2008 data. The $\btopik$
amplitudes are written in terms of four independent diagrams
and the phase $\gamma$.  In addition, in $S_{\sss CP}$ the
phase of $\bd$-$\bdbar$ mixing ($\beta$) appears. Our fits
therefore involve 9 theoretical parameters: the magnitudes of
$P'_{tc}$, $T'$, $C'$, $P'_{uc}$, three relative strong
phases, and the weak phases $\beta$ and $\gamma$. Note: for
$\beta$, we add the additional constraint of $\beta =
(21.66^{+0.95}_{-0.87})^\circ$ \cite{CKMfitter}, which is
obtained mainly from the measurement of CP violation in
$\bd(t)\to J/\Psi \ks$ and other ${\bar b} \to {\bar c} c
{\bar s}$ decays. Also, whenever experimental data show
asymmetrical errors, we take the larger one as the 1$\sigma$
standard deviation in the fits.

We first fit to only the $\btopik$ data of
Table~\ref{tab:data}. Before presenting the results of this
fit, we note the following.  In general, several solutions
will be found when a fit is done, many of which will indicate
new physics (NP).  How do we know that one of these is not
the correct solution?  We can never be absolutely sure, but
in order to make progress, we have to make some plausible
assumptions.  In particular, we assume that the NP, if
present, is not enormous.  If it were, it probably would
already have been seen elsewhere.  In the present case, there
is one solution which has a better value of
$\chi^2_{min}/d.o.f.$ than the one eventually chosen.  It is
shown in Table~\ref{neglectedfits}. In the SM,
$|P'_{uc}/P'_{tc}|$ is estimated to be in the range
${\bar\lambda}^3$-${\bar\lambda}^2$ (${\bar\lambda} \sim
0.2$).  However, in the solution of
Table~\ref{neglectedfits}, it is found that $|P'_{uc}| >
|P'_{tc}|$.  In order for this to occur, NP must be present,
and it must be extremely large.  This is counter to our
assumption, and we therefore discard this solution on the
basis that it cannot represent the true situation.

\begin{table}[tbh]
\center
\begin{tabular}{ccccc}
\hline
\hline
$\chi^2_{min}/d.o.f.$ & $|P'_{tc}|$ & $|T'|$ & $|C'|$ &
$|P'_{uc}|$ \\
\hline
$0.02/1$ & $25.1 \pm 3.6$ & $11.6 \pm 2.1$ &
$9.1 \pm 2.4$ & $42.9 \pm 3.2$ \\
\hline
\hline
\end{tabular}
\caption{Discarded solution of the fit to $P'_{tc}$, $T'$,
  $C'$, $P'_{uc}$, $\beta$ and $\gamma$ in the SM.  The fit
  includes the constraint $\beta =
  (21.66^{+0.95}_{-0.87})^\circ$. The amplitude is in units
  of eV.}
\label{neglectedfits}
\end{table}

The solution which is retained is shown in
Table~\ref{resultsfit1}. Note: throughout this paper, we
adopt the convention that the strong phase of $P'_{tc}$ is
zero.  Now, the underlying diagram $P'_{tc}$ does have a
strong phase, coming from rescattering of the ${\bar b} \to
{\bar c} c {\bar s}$ tree diagram\footnote{This is theory
input.  Experimentally, one can only measure differences of
strong phases.  However, there is a theoretical framework
describing the generation of individual strong phases.  This
framework of rescattering is used in numerous papers, and is
included in all theories of QCD.  We use it throughout this
paper.}. So we have shifted the strong phases of all diagrams
by this quantity.  Thus, when we write $\delta_{\sss T'}$, it
really corresponds to the strong phase of the underlying $T'$
diagram minus $\delta_{\sss P'_{tc}}$, and similarly for
$\delta_{\sss C'}$ and $\delta_{\sss P'_{uc}}$ (and
$\delta_{\sss NP}$ later in the paper).

We find that $\chi^2_{min}/d.o.f. = 0.52/1~(47\%)$. (The
number in parentheses indicates the quality of the fit, and
depends on $\chi^2_{min}$ and $d.o.f.$ individually. 50\% or
more is a good fit; fits which are substantially less than
50\% are poorer.) Here the quality of the fit is
fair. However, there is clearly a possible hint of NP.  The
most important point is that the value of $\gamma$ extracted
is very far from that obtained from independent measurements,
$\gamma \sim 67^\circ$. This leads to a discrepancy of
$3.5\sigma$. (It is also true that the central value of
$|P'_{uc}|$ is very large.  But the error is also large, so
this is not as much of a problem.)

\begin{table}[tbh]
\center
\begin{tabular}{ccccc}
\hline
\hline
$\chi^2_{min}/d.o.f.$ & $|P'_{tc}|$ & $|T'|$ & $|C'|$ &
$|P'_{uc}|$ \\
\hline
$0.52/1$ & $67.7 \pm 11.8$ & $19.6 \pm 6.9$ & $14.9 \pm 6.6$
& $20.5 \pm 13.3$ \\
\hline
$\delta_{\sss T'}$ & $\delta_{\sss C'}$ & $\delta_{\sss
P'_{uc}}$ & $\beta$ & $\gamma$ \\
\hline
$(6.0 \pm 4.0)^\circ$ & $(-11.7 \pm 6.8)^\circ$ & $(-0.7 \pm
2.3)^\circ$ & $(21.66 \pm 0.95)^\circ$ & $(35.3 \pm
7.1)^\circ$\\
\hline
\hline
\end{tabular}
\caption{Results of the fit to $P'_{tc}$, $T'$, $C'$,
  $P'_{uc}$, $\beta$ and $\gamma$ in the SM.  The fit
  includes the constraint $\beta =
  (21.66^{+0.95}_{-0.87})^\circ$.  The amplitude is in units
  of eV.}
\label{resultsfit1}
\end{table}

Since the quality of fit is only fair, it is difficult to
conclude definitively that NP is indicated. We therefore
perform a second fit, in which an additional constraint on
$\gamma$ from independent measurements has been
imposed\footnote{There are no $\btos$ measurements entering
this value of $\gamma$, so there is no inconsistency in using
it in the search for $\btos$ NP in the $\btopik$ puzzle.}:
$\gamma = (66.8^{+5.4}_{-3.8})^\circ$ \cite{CKMfitter}. This
is the ``SM fit.'' This fit may not be good, but the question
is: how bad is it?  Could the SM still be considered to
explain the $\btopik$ data?

There is one discarded solution; that which is kept is shown
in Table~\ref{resultsfit2}. We find that
$\chi^2_{min}/d.o.f. = 3.2/2~(20\%)$.  In addition to the
poor quality of fit, the ratio $|T'/P'_{tc}|$ is on the small
side.  Neither of these deficiencies definitively excludes
the SM as the explanation of the $\btopik$ data.  However,
they do negatively impact the SM fit.  Nevertheless, one
might consider that the $\btopik$ measurements are not
actually at odds with the SM. One therefore sees that the
$\btopik$ puzzle is now in an uncertain situation -- it is
not clear if NP is indicated, or if the SM is sufficient.

\begin{table}[tbh]
\center
\begin{tabular}{ccccc}
\hline
\hline
$\chi^2_{min}/d.o.f.$ & $|P'_{tc}|$ & $|T'|$ & $|C'|$ &
$|P'_{uc}|$ \\
\hline
$3.2/2$ & $50.5 \pm 1.8$ & $5.9 \pm 1.8$ & $3.4 \pm 1.0$ &
$2.3 \pm 4.9$ \\
\hline
$\delta_{\sss T'}$ & $\delta_{\sss C'}$ & $\delta_{\sss
P'_{uc}}$ & $\beta$ & $\gamma$ \\
\hline 
$(25.5 \pm 11.2)^\circ$ & $(252.0 \pm 36.1)^\circ$ & $(-9.9
\pm 27.8)^\circ$ & $(21.65 \pm 0.95)^\circ$ & $(66.5 \pm
5.5)^\circ$ \\
\hline 
\hline
\end{tabular}
\caption{Results of the fit to $P'_{tc}$, $T'$, $C'$,
  $P'_{uc}$, $\beta$ and $\gamma$ in the SM.  The fit
  includes the constraints $\beta =
  (21.66^{+0.95}_{-0.87})^\circ$ and $\gamma =
  (66.8^{+5.4}_{-3.8})^\circ$. The amplitude is in units of
  eV.}
\label{resultsfit2}
\end{table}

Recent analyses have examined new contributions to $\btos$
transitions \cite{newbtos}.  When looking at $\btopik$
decays, they have focused on the difference between $S_{\sss
CP}(\pi^0 K^0)$ and the measured indirect CP asymmetry in
$\bd\to J/\Psi\ks$.  However, we find that there is little
difficulty in reproducing the measured value of $S_{\sss
CP}(\pi^0 K^0)$ in the fits.  (Also, the fact that there is a
greater than $5\sigma$ difference between $A_{\sss CP}(\pi^0
K^+)$ and $A_{\sss CP}(\pi^- K^+)$ \cite{BL} is not a problem
once the smaller diagrams $\pewp$, $C'$ and $\pewcp$ are
taken into account \cite{thanks}.) Rather, it is the direct
CP asymmetry in $\bd \to \pi^0 K^0$ which causes the most
problems.  This is illustrated in Table~\ref{pulls}, in which
the predictions for each of the observables are given for the
fits of Table~\ref{resultsfit1} (Fit 1) and
Table~\ref{resultsfit2} (Fit 2), along with the ``pull'' from
the data.  (The pull is defined as (data central value $-$
theory prediction) / (data error).) We see that $A_{\sss
CP}(\pi^0 K^0)$ has the largest pull in both fits (and that
$S_{\sss CP}(\pi^0 K^0)$ has a small pull).

\begin{table}[tbh]
\center
\begin{tabular}{ccc}
\hline\hline
Obs. & Fit 1 & Fit 2 \\
\hline
$BR(\pi^+ K^0)$ & $23.1 \; (+0.02)$ & $23.7 \; (-0.57)$ \\
$A_{\sss CP}(\pi^+ K^0)$ & $0.014 \; (-0.21)$ & $0.016 \; (-0.29)$ \\
$BR(\pi^0 K^+)$ & $12.9 \; (-0.03)$ & $12.5 \; (+0.72)$ \\
$A_{\sss CP}(\pi^0 K^+)$ & $0.05 \; (+0.15)$ & $0.04 \; (+0.27)$ \\
$BR(\pi^- K^+)$ & $19.4 \; (+0.05)$ & $19.7 \; (-0.46)$ \\
$A_{\sss CP}(\pi^- K^+)$ & $-0.098 \; (-0.04)$ & $-0.097 \; (-0.12)$ \\
$BR(\pi^0 K^0)$ & $9.8 \; (-0.07)$ & $9.3 \; (+0.88)$ \\
$A_{\sss CP}(\pi^0 K^0)$ & $-0.08 \; (+0.66)$ & $-0.12 \; (+1.10)$ \\
$S_{\sss CP}(\pi^0 K^0)$ & $0.58 \; (-0.03)$ & $0.58 \; (-0.08)$ \\
\hline\hline
\end{tabular}
\caption{Predictions of the $\btopik$ decay observables based
upon the best-fitted results in Table~\ref{resultsfit1} (Fit
1) and Table~\ref{resultsfit2} (Fit 2).  Branching ratios are
given in units of $10^{-6}$. Numbers in parentheses are the
corresponding pulls.}
\label{pulls}
\end{table}

There is a slight possible complication, related to the
$A_{\sss CP}(\pi^0 K^0)$ measurement by the BaBar
Collaboration ($-0.13 \pm 0.13$) and the Belle Collaboration
($0.14 \pm 0.14$) -- the central values do not quite agree
with each other.  (The average is nearly zero, which is used
in our previous fits.) However, both of our predictions in
Table~\ref{pulls} favor the BaBar measurement.  We therefore
also consider the above-mentioned two fits using the BaBar
value for $A_{\sss CP}(\pi^0 K^0)$.  The results are shown in
Table~\ref{babarfit}.  Indeed, we observe an improved quality
of fit (from $47\%$ and $20\%$ to $76\%$ and $43\%$,
respectively).  On the other hand, despite this improvement,
there are similar puzzling features as found before. For
example, Fit $1'$ still gives a value of $\gamma$ about $3.5
\sigma$ below other determinations of this quantity.  And
even though the ratio $|T'/P'_{tc}|$ in Fit $2'$ is found to
be larger than that in Fit 2, a new problem arises in that
$|C'/T'| = 1.6 \pm 0.6$.  We note in passing (without
presenting all the details) that if we take instead the Belle
measurement for $A_{\sss CP}(\pi^0 K^0)$ in the two fits, the
fit quality drops to $18\%$ and $7\%$, respectively.  The
results corresponding to Fit 1 have the problems that $\gamma
= 96.4 \pm 12.4$, about $2.2 \sigma$ above the other
measurements, and $|P'_{uc}| = 31.9 \pm 5.4$ is large.  The
problem in the results corresponding to Fit 2 is still the
small $|T'/P'_{tc}|$ ratio.

\begin{table}[tbh]
\center
\begin{tabular}{ccccc}
\hline\hline
$\chi^2_{min}/d.o.f.$ & $|P'_{tc}|$ & $|T'|$ & $|C'|$ &
$|P'_{uc}|$ \\ 
\hline
Fit $1'$: $0.095/1$ & $66.9 \pm 11.9$ & $18.8 \pm 7.3$ &
$14.1 \pm 7.0$ & $19.9 \pm 13.3$ \\
\hline
$\delta_{\sss T'}$ & $\delta_{\sss C'}$ & $\delta_{\sss
P'_{uc}}$ & $\beta$ & $\gamma$ \\
\hline 
$(5.4 \pm 4.3)^\circ$ & $(-13.6 \pm 8.4)^\circ$ & $(0.0 \pm
2.4)^\circ$ & $(21.66 \pm 0.95)^\circ$ & $(35.9 \pm
7.7)^\circ$ \\
\hline\hline
$\chi^2_{min}/d.o.f.$ & $|P'_{tc}|$ & $|T'|$ & $|C'|$ &
$|P'_{uc}|$ \\
\hline
Fit $2'$: $1.7/2$ & $41.5 \pm 2.4$ & $8.5 \pm 2.5$ & $13.7
\pm 2.6$ & $10.7 \pm 3.9$ \\
\hline
$\delta_{\sss T'}$ & $\delta_{\sss C'}$ & $\delta_{\sss
P'_{uc}}$ & $\beta$ & $\gamma$ \\
\hline 
$(139.2 \pm 12.8)^\circ$ & $(212.2 \pm 7.2)^\circ$ & $(184.0
\pm 4.3)^\circ$ & $(21.59 \pm 0.95)^\circ$ & $(64.6 \pm
4.9)^\circ$ \\
\hline
\hline
\end{tabular}
\caption{Results of the fits to $P'_{tc}$, $T'$, $C'$,
  $P'_{uc}$, $\beta$ and $\gamma$ in the SM. The averaged
  experimental data given in Table~\ref{tab:data} is used,
  except that only the BaBar measurement of $A_{\sss
  CP}(\pi^0 K^0) = -0.13 \pm 0.13$ is taken. The fits include
  the constraints $\beta = (21.66^{+0.95}_{-0.87})^\circ$.
  For $\gamma$, we impose no constraint (Fit $1'$), or
  $\gamma = (66.8^{+5.4}_{-3.8})^\circ$ (Fit $2'$). The
  amplitude is in units of eV.}
\label{babarfit}
\end{table}

Assuming that new physics is present in $\btopik$, it is now
important to know what type of NP it is.  We follow the
approach developed in Ref.~\cite{DLNP}. All NP operators in
$\btopik$ decays take the form ${\cal O}_{\sss NP}^{ij,q}
\sim {\bar s} \Gamma_i b \, {\bar q} \Gamma_j q$ ($q = u,d$),
where $\Gamma_{i,j}$ represent Lorentz structures, and color
indices are suppressed. These operators contribute through
the matrix elements $\bra{\pi K} {\cal O}_{\sss NP}^{ij,q}
\ket{B}$. In general, each matrix element has its own NP weak
and strong phases.

In Ref.~\cite{DLNP}, it is argued that all NP strong phases
are negligible.  The reason is that the strong phase of the
SM diagram $P'_{tc}$ is generated by rescattering of the
${\bar b} \to {\bar c} c {\bar s}$ tree diagram, which is
about 100 times as big.  On the other hand, the NP strong
phase can only be generated by rescattering of the NP
diagram itself, i.e.\ self-rescattering.  It is therefore
negligible compared to the strong phase of $P'_{tc}$. In this
case, one can combine all NP matrix elements into a single NP
amplitude, with a single weak phase:
\beq
\sum \bra{\pi K} {\cal O}_{\sss NP}^{ij,q} \ket{B} = \ANPq
e^{i \Phi_q} ~.
\eeq
There are two classes of NP amplitudes, differing only in
their color structure: ${\bar s}_\alpha \Gamma_i b_\alpha \,
{\bar q}_\beta \Gamma_j q_\beta$ and ${\bar s}_\alpha
\Gamma_i b_\beta \, {\bar q}_\beta \Gamma_j q_\alpha$ ($q =
u,d$). They are denoted $\ApNPqph$ and $\ApNPCqph$,
respectively \cite{BNPmethods}. Here, $\Phi'_q$ and
$\Phi_q^{\prime {\sss C}}$ are the NP weak phases; the strong
phases are zero. Each of these contributes differently to the
various $\btopik$ decays. In general, ${\cal A}^{\prime,q}
\ne {\cal A}^{\prime {\sss C}, q}$ and $\Phi'_q \ne
\Phi_q^{\prime {\sss C}}$. Note: despite the
``color-suppressed'' index $C$, the matrix elements
$\ApNPCqph$ are not necessarily smaller than $\ApNPqph$.

There are three NP matrix elements which contribute to the
$\btopik$ amplitudes: $\ApNPcomb \equiv - \ApNPuph +
\ApNPdph$, $\ApNPCuph$, and $\ApNPCdph$ \cite{BNPmethods}.
The first operator corresponds to including NP only in the
color-favored electroweak penguin amplitude: $\ApNPcomb
\equiv -\pewpnp \, e^{i \Phi'_{\sss EW}}$. Nonzero values of
$\ApNPCuph$ and/or $\ApNPCdph$ imply the inclusion of NP in
both the gluonic and color-suppressed electroweak penguin
amplitudes, $P'_{\sss NP} \, e^{i \Phi'_{\sss P}}$ and
$\pewcpnp \, e^{i \Phi^{\prime {\sss C}}_{\sss EW}}$,
respectively~\cite{Baek:2006ti}:
\bea
P'_{\sss NP} \, e^{i \Phi'_{\sss P}} & \equiv & \frac13
\ApNPCuph\ + \frac23 \ApNPCdph~, \nn\\
\pewcpnp \, e^{i \Phi^{\prime {\sss C}}_{\sss EW}} & \equiv &
\ApNPCuph\ - \ApNPCdph ~.
\eea
(In Ref.~\cite{Baeketal}, NP only in the gluonic penguin
amplitude was referred to as ``isospin-conserving NP:''
$\ApNPCuph\ = \ApNPCdph$, $\ApNPcomb = 0$.)

The $\btopik$ amplitudes can now be written in terms of the
SM amplitudes, along with the NP matrix elements.  We neglect
only the (small) SM diagram $P'_{uc}$:
\bea
\label{BpiKNPamps}
A^{+0} &\!\!=\!\!& -P'_{tc} -\frac13 \pewcp + P'_{\sss NP} \,
e^{i \Phi'_{\sss P}} -\frac13 \pewcpnp \, e^{i \Phi^{\prime
{\sss C}}_{\sss EW}} ~, \nn\\
\sqrt{2} A^{0+} &\!\!=\!\!& P'_{tc} - T' \, e^{i\gamma} -
\pewp -C' \, e^{i\gamma} -\frac23 \pewcp \nn\\
& & \hskip1truecm -~\pewpnp \, e^{i \Phi'_{\sss EW}} -
  P'_{\sss NP} \, e^{i \Phi'_{\sss P}} - \frac23 \pewcpnp \,
  e^{i \Phi^{\prime {\sss C}}_{\sss EW}} ~, \nn\\
A^{-+} &\!\!=\!\!& P'_{tc} - T' \, e^{i\gamma} -\frac23
  \pewcp - P'_{\sss NP} \, e^{i \Phi'_{\sss P}} - \frac23
  \pewcpnp \, e^{i \Phi^{\prime {\sss C}}_{\sss EW}} ~, \nn\\
\sqrt{2} A^{00} &\!\!=\!\!& -P'_{tc} - \pewp -C' \,
e^{i\gamma} -\frac13 \pewcp \nn\\
& & \hskip1truecm -~\pewpnp \, e^{i \Phi'_{\sss EW}} +
  P'_{\sss NP} \, e^{i \Phi'_{\sss P}} - \frac13 \pewcpnp \,
  e^{i \Phi^{\prime {\sss C}}_{\sss EW}} ~.
\eea

Even if the value of $\gamma$ is taken from independent
measurements, there are too many theoretical parameters to
perform a fit containing all three NP operators. It is
therefore necessary to make some theoretical assumptions.  As
in Ref.~\cite{Baeketal}, we assume that a single NP amplitude
dominates, and consider $P'_{\sss NP} \, e^{i \Phi'_{\sss
P}}$, $\pewpnp \, e^{i \Phi'_{\sss EW}}$, and $\pewcpnp \,
e^{i \Phi^{\prime {\sss C}}_{\sss EW}}$ individually.

However, we have not included all the information at our
disposal. The strong phase of the $T'$ diagram can arise only
from self-rescattering. Thus, like the NP amplitudes, this
strong phase is expected to be very small.  We take this into
account by adding the constraint $\delta_{\sss T'} =
\delta_{\sss NP}$.

The results of the NP fits are given in
Table~\ref{resultsNPfit1}.  In the first fit, $P'_{\sss NP}
\, e^{i \Phi'_{\sss P}}$ is assumed to be nonzero.  Several
of the entries in the Table are given as NA.  This stands for
``not applicable,'' and means that there are no limits on the
corresponding theoretical parameters.  This can be understood
as follows.  This type of NP always appears in the following
combination in the $\btopik$ amplitudes: $P'_{tc} - P'_{\sss
NP} \ e^{i \Phi'_{\sss P}}$.  This contains the 4 quantities
$|P'_{tc}|$, $|P'_{\sss NP}|$, $\delta_{\sss NP}$\footnote{In
our convention, $\delta_{\sss NP} = \delta_{new~phys} -
\delta_{\sss P'_{tc}}$.  However, the new-physics strong
phase $\delta_{new~phys}$ is negligible, so that
$\delta_{\sss NP} = - \delta_{\sss P'_{tc}}$.}, and
$\Phi'_{\sss P}$.  However, here these are not all
independent.  The easiest way to see this is to use the
convention in which $e^{i \delta_{\sss P'_{tc}}}$ multiplies
$|P'_{tc}|$.  For $B$ and ${\bar B}$ decays, the combinations
are:
\bea
{\tilde P}^{\prime}_{tc} e^{i \delta_{\sss P'_{tc}}} -
{\tilde P}^{\prime}_{\sss NP} \ e^{i \Phi'_{\sss P}} & \equiv
& z ~, \nn\\
{\tilde P}^{\prime}_{tc} e^{i \delta_{\sss P'_{tc}}} -
{\tilde P}^{\prime}_{\sss NP} \ e^{-i \Phi'_{\sss P}} &
\equiv & z' ~.
\eea
Here, the diagrams are written with a tilde to indicate the
different convention, and that the strong phases are given
explicitly. $z$ and $z'$ are complex numbers; their 4 real
and imaginary parts can be written in terms of the 4
theoretical parameters.  However, it is clear from the above
expressions that ${\rm Re} \, z = {\rm Re} \, z'$.  The 4
parameters $|P'_{tc}|$, $|P'_{\sss NP}|$, $\delta_{\sss NP}$
and $\Phi'_{\sss P}$ are therefore not independent, and so
their allowed ranges cannot be fixed; they are given as NA in
Table~\ref{resultsNPfit1}. $\delta_{\sss C'}$ only appears in
$\btopik$ observables in tandem with other strong phases.
These are NA, so that $\delta_{\sss C'}$ is as well.  Since
there are no constraints on the NP parameters, this fit
[$\chi^2_{min}/d.o.f.\ = 3.6/2 \, (17\%)$] is basically that
of the SM.  Indeed, this result is quite similar to that
given in Table~\ref{resultsfit2} (the small differences are
due to the fact that $P'_{uc}$ is neglected here).

\begin{table}[tbh]
\center
\begin{tabular}{cccc}
\hline
\hline
$\chi^2_{min}/d.o.f.$ & $|P'_{tc}|$ & $|T'|$ & $|C'|$ \\
\hline
$3.6/2$ & NA & $5.9 \pm 2.0$ & $3.6 \pm 1.0$ \\
\hline
$|P'_{\sss NP}|$ & $\delta_{\sss C'}$ & $\delta_{\sss NP}$ &
$\Phi'_{\sss P}$ \\
\hline 
NA & NA & NA & NA \\
\hline 
\hline
$\chi^2_{min}/d.o.f.$ & $|P'_{tc}|$ & $|T'|$ & $|C'|$ \\
\hline
$0.4/2$ & $48.2 \pm 1.3$ & $2.6 \pm 0.4$ & $16.1 \pm 28.4$ \\
\hline
$|P'_{\sss EW, NP}|$ & $\delta_{\sss
C'}$ & $\delta_{\sss NP}$ & $\Phi'_{\sss EW}$ \\
\hline 
$20.1 \pm 22.3$ & $(254.8 \pm 21.8)^\circ$ & $(95.4 \pm
9.6)^\circ$ & $(37.6 \pm 51.8)^\circ$ \\
\hline 
\hline
$\chi^2_{min}/d.o.f.$ & $|P'_{tc}|$ & $|T'|$ & $|C'|$ \\
\hline
$2.5/2$ & $48.2 \pm 1.3$ & $1.9 \pm 1.4$ & $9.4 \pm 2.3$ \\
\hline
$|P_{\sss EW, NP}^{\prime\sss C}|$ & $\delta_{\sss C'}$ &
$\delta_{\sss NP}$ & $\Phi^{\prime {\sss C}}_{\sss EW}$ \\
\hline 
$16.5 \pm 15.2$ & $(192.4 \pm 12.3)^\circ$ & $(97.8 \pm
15.3)^\circ$ & $(183.9 \pm 7.8)^\circ$ \\
\hline 
\hline
\end{tabular}
\caption{Results of the fits to $P'_{tc}$, $T'$, $C'$,
  $\beta$, $\gamma$ and a NP amplitude.  The fits include the
  constraints $\beta = (21.66^{+0.95}_{-0.87})^\circ$ and
  $\gamma = (66.8^{+5.4}_{-3.8})^\circ$, and in all cases,
  the best-fit values of $\beta$ and $\gamma$ are consistent
  with these. The constraint $ \delta_{\sss T'} =
  \delta_{\sss NP}$ is also added. The amplitude is in units
  of eV. The entry NA (not applicable) is explained in the
  text.}
\label{resultsNPfit1}
\end{table}

This shows that global fits are insensitive to NP in the
gluonic penguin amplitude.  If one wishes to investigate the
possibility that such NP can account for the $\btopik$ data,
it is necessary to consider specific models and perform a
model-dependent calculation.  Note: in previous analyses, the
NP operators $\ApNPCuph$ and $\ApNPCdph$ were considered, and
constraints on the theoretical parameters given. {}From the
above, we see that these constraints are due solely to NP in
the color-suppressed electroweak penguin.

In the second fit, $\pewpnp \, e^{i \Phi'_{\sss EW}}$ is
assumed to be nonzero.  In this case, the fit is excellent:
$\chi^2_{min}/d.o.f.\ = 0.4/2 \, (82\%)$.  However, the
central value of $|C'/T'|$ is enormous, far outside of the
allowed range [Eq.~(\ref{ampsizes})].  Even though the errors
are large, this is worrisome.  The third fit takes $\pewcpnp
\, e^{i \Phi^{\prime {\sss C}}_{\sss EW}}$ to be nonzero.
Here the fit is poor -- $\chi^2_{min}/d.o.f.\ = 2.5/2 \,
(28\%)$ -- but might still be considered as viable.  On the
other hand, even here $|C'/T'|$ is far too large, which poses
problems for this scenario.

Because the value of $|C'/T'|$ is large in two of the fits,
we redo the fits with the constraint $|C'/T'| = 0.5$. The
results are given in Table~\ref{resultsNPfit2}.  As before,
the fit with $P'_{\sss NP} \, e^{i \Phi'_{\sss P}}$ just
reproduces that of the SM.  The improved quality of fit --
$\chi^2_{min}/d.o.f.\ = 3.7/3 \, (29\%)$ -- simply
corresponds to the fact that the d.o.f.\ has increased from 2
to 3.  In the second fit, with $\pewpnp \, e^{i \Phi'_{\sss
EW}} \ne 0$, the quality of fit has decreased markedly:
$\chi^2_{min}/d.o.f.\ = 3.0/3 \, (39\%)$.  While this is
still better than the SM, the improvement with NP is hardly
convincing (and the fit is at best fair).  The third fit has
a nonzero $\pewcpnp \, e^{i \Phi^{\prime {\sss C}}_{\sss
EW}}$.  It is not particularly good: $\chi^2_{min}/d.o.f.\ =
3.8/3 \, (28\%)$.  In it, the value of the NP parameter is
rather small, so that this scenario is also essentially that
of the SM.

\begin{table}[tbh]
\center
\begin{tabular}{cccc}
\hline
\hline
$\chi^2_{min}/d.o.f.$ & $|P'_{tc}|$ & $|T'|$ & $|P'_{\sss
NP}|$ \\
\hline
$3.7/3$ & NA & $6.6 \pm 1.1$ & NA \\
\hline
$\delta_{\sss C'}$ & $\delta_{\sss NP}$ &
$\Phi'_{\sss P}$ & \\
\hline 
NA & NA & NA & \\
\hline 
\hline
$\chi^2_{min}/d.o.f.$ & $|P'_{tc}|$ & $|T'|$ & $|P'_{\sss EW,
  NP}|$ \\
\hline
$3.0/3$ & $48.0 \pm 0.6$ & $2.6 \pm 0.3$ & $15.7 \pm 3.6$ \\

\hline
$\delta_{\sss
C'}$ & $\delta_{\sss NP}$ & $\Phi'_{\sss EW}$ & \\
\hline 
$(182.5 \pm 53.1)^\circ$ & $(98.4 \pm
4.7)^\circ$ & $(-11.6 \pm 5.7)^\circ$ & \\
\hline 
\hline
$\chi^2_{min}/d.o.f.$ & $|P'_{tc}|$ & $|T'|$ & $|P_{\sss EW,
NP}^{\prime\sss C}|$ \\
\hline
$3.8/3$ & $49.8 \pm 0.7$ & $6.5 \pm 1.4$ & $2.1 \pm 6.2$ \\
\hline
$\delta_{\sss C'}$ &
$\delta_{\sss NP}$ & $\Phi^{\prime {\sss C}}_{\sss EW}$ & \\
\hline 
$(274.7 \pm 59.2)^\circ$ & $(15.6 \pm
10.8)^\circ$ & $(69.7 \pm 67.4)^\circ$ & \\
\hline 
\hline
\end{tabular}
\caption{Results of the fits to $P'_{tc}$, $T'$, $C'$,
  $\beta$, $\gamma$ and a NP amplitude.  The fits include the
  constraints $\beta = (21.66^{+0.95}_{-0.87})^\circ$ and
  $\gamma = (66.8^{+5.4}_{-3.8})^\circ$, and in all cases,
  the best-fit values of $\beta$ and $\gamma$ are consistent
  with these. The constraints $ \delta_{\sss T'} =
  \delta_{\sss NP}$ and $|C'/T'| = 0.5$ are also added. The
  amplitude is in units of eV. The entry NA (not applicable)
  is explained in the text.}
\label{resultsNPfit2}
\end{table}

If there is NP, the $\btopik$ data point to the
color-favored electroweak penguin amplitude.  This is the
same situation as in previous analyses.  However, whereas in
the past, this was a 3-4$\sigma$ effect, now it is much less
clear.  Both the SM and NP fits are only fair, and the NP
case is an improvement on the SM only by a small amount.

In summary, we have performed an update of the $\btopik$
puzzle by performing several fits comparing the $\btopik$
measurements as of 2008 with the predictions of the Standard
Model (SM).  Our first fit involves only the $\btopik$ data.
We find a possible hint of new physics (NP), since the
extracted value of the weak phase $\gamma$ disagrees with
that of independent measurements.  However, the fit is only
fair.  We then constrain $\gamma$ to satisfy these
measurements and perform another fit (the SM fit).  We find
that the fit is on the poor side.  However, it is not so bad
that it is excluded.  We now have the situation where the
$\btopik$ data somewhat favor NP over the SM, but not by a
huge amount.

We now assume that NP is present.  It can appear in the
gluonic penguin, the color-favored electroweak penguin,
and/or the color-suppressed electroweak penguin.  We consider
each of these NP scenarios individually.  We first show that
the global fit is insensitive to NP in the gluonic penguin --
with this type of NP, one simply reproduces the result of the
SM.  If NP is in the color-suppressed electroweak penguin,
the best fit has a small contribution of NP.  Thus, this is
also very much like the SM.  The only case in which one finds
a large, nonzero NP operator is when it contributes to the
color-favored electroweak penguin amplitude.  This appears to
be as in previous analyses.  The difference is that now this
fit is only fair, and it is better than that of the SM by
only a small amount.

The conclusion is that, while the $\btopik$ puzzle is still
present, it is considerably weaker.  Neither the SM nor NP
gives an excellent fit to the data.  And while the $\btopik$
measurements do point to NP in the color-favored electroweak
penguin, this is not a clear indication, as it was before,
and the NP scenario is only a little better than that of the
SM.

\bigskip
\noindent
{\bf Acknowledgments}:
%\bigskip
D.L. would like to thank A. Datta and M. Imbeault for helpful
conversations.  The authors also thank H.-n.\ Li for useful
comments on the manuscript. This work was financially
supported by by the Korea Research Foundation Grant funded by
the Korean Government (MOEHRD) No. KRF-2007-359-C00009 (SB),
by Grant No.~NSC~97-2112-M-008-002-MY3 of Taiwan and the
hospitality of NCTS-Hsinchu (CC), and by NSERC of Canada
(DL).

%%%%%%%%%%%%%%%%%%%%% REFERENCES %%%%%%%%%%%%%%%%%%%%%

\end{document}